\documentclass[fleqn,usenatbib,times]{mnras}

\usepackage{newtxtext,newtxmath}
\usepackage{graphicx}
\usepackage{tikz}
\usepackage{wasysym}
\usepackage{multirow}
\usepackage{hyperref}
\def\equationautorefname~#1\null{(#1)\null}
\usepackage[T1]{fontenc}
\usepackage{ae,aecompl}

\usepackage[nameinlink,capitalize]{cleveref}
\crefformat{section}{\S#2#1#3} 
\crefformat{subsection}{\S#2#1#3}
\crefformat{subsubsection}{\S#2#1#3}

\usepackage{graphicx}	
\usepackage{amsmath}	
\usepackage{soul}

\title[Density discontinuities in planetary SPH]{Dealing with density discontinuities in planetary SPH simulations}

\author[S. Ruiz-Bonilla et al.]{S. Ruiz-Bonilla$^{1,2}$\thanks{E-mail: sergio.ruiz-bonilla@durham.ac.uk},
J. Borrow$^{3},$
V. R. Eke$^{1},$
J. A. Kegerreis$^{1,4},$
R. J. Massey$^{1},$
\newauthor
T. D. Sandnes$^{1},$
L. F. A. Teodoro$^{5,6}$
\\
$^{1}$Institute for Computational Cosmology, Durham University, South Road, Durham DH1 3LE, UK\\
$^{2}$Institute for Data Science, Durham University, South Road, Durham DH1 3LE, UK \\
$^{3}$Department of Physics, Kavli Institute for Astrophysics and Space Research, Massachusetts Institute of Technology, Cambridge, MA 02139, USA \\
$^{4}$NASA Ames Research Center, Moffett Field, CA, USA\\
$^{5}$BAERI/NASA Ames Research Center, Moffett Field, CA, USA\\
$^{6}$School of Physics and Astronomy, University of Glasgow, G12 8QQ, Scotland, UK
}

\date{Accepted XXX. Received YYY; in original form ZZZ}

\pubyear{2021}

\begin{document}
\label{firstpage}
\pagerange{\pageref{firstpage}--\pageref{lastpage}}
\maketitle

\begin{abstract}
Density discontinuities cannot be precisely modelled in standard formulations of smoothed particles hydrodynamics (SPH) because the density field is defined smoothly as a kernel-weighted sum of neighbouring particle masses.
This is a problem when performing simulations of giant impacts between proto-planets, for example, because planets typically do have density discontinuities both at their surfaces and at any internal boundaries between different materials.
The inappropriate densities in these regions create artificial forces that effectively suppress mixing between particles of different material and,
as a consequence, this problem introduces a key unknown systematic error into studies that rely on SPH simulations.
In this work we present a novel, computationally cheap method that deals simultaneously with both of these types of density discontinuity in SPH simulations.
We perform standard hydrodynamical tests and several example giant impact simulations, and compare the results with standard SPH.
In a simulated Moon-forming impact using $10^7$ particles, the improved treatment at boundaries affects at least 30$\%$ of the particles at some point during the simulation.

\end{abstract}

\begin{keywords}
methods: numerical -- hydrodynamics -- planets and satellites: formation 
\end{keywords}


\section{Introduction}
\label{introduction}
A key chapter in the solar system's history involves impacts between planet-sized objects.
This giant impact phase of planet and satellite formation is responsible for many of the features we see today in our solar system.
To name a few: the formation of our Moon \citep[e.g.][]{Hartmann+1975, cameron1976, Benz+1987}, the tilted spin axis of Uranus \citep[e.g.][]{Slattery+1992, Kegerreis+2018, Reinhardt2019}, the formation of the Pluto-Charon system \citep[e.g.][]{McKinnon1984, McKinnon1989, Canup2005}, the Mars hemispheric dichotomy \citep[e.g.][]{Wilhelms1984}, or the origin of Mercury's high core:mantle ratio \citep[e.g.][]{Benz1988, Chau2018}.

An ideal tool for studying giant impacts is smoothed particle hydrodynamics (SPH). SPH is a particle-based method used in a wide range of astrophysical and engineering topics \citep{Springel2010, Monaghan2012}.
It is the most commonly used option for studying giant impacts because of the complexity and anisotropy of these highly non-linear interactions. Compared with grid-based hydrodynamical codes, SPH has the advantages of naturally following the provenance of material and being readily combined with efficient gravity solvers.

Despite its many positive points, the hydrodynamical part of SPH can still have difficulties treating the mixing of particles that represent different materials.
In the standard density-energy formulation of SPH, density discontinuities cannot be accurately represented because of the smoothing inherent in the definition of the density field. However, density profiles of differentiated planets do have discontinuities, typically both between different material layers (e.g. core to mantle boundary) and the outer surface. 
The standard SPH formulation creates artificial forces that act like an effective surface tension at these discontinuities, repelling one material from the other and suppressing mixing between different materials.
While the cause of this numerical artefact is clear, the consequences for the mixing of materials during giant impact simulations are rarely considered \citep{Deng2019ApJ...870..127D,Deng2019ApJ...887..211D}. As such, this represents a significant and unquantified systematic uncertainty for standard simulations. This numerical issue could be crucial in the modelling of many giant impact problems, with examples being how much mixing would have been provoked in the core of Jupiter by a giant impact \citep{Liu2019}, and what the distribution of iron is in the debris of the hypothesised Moon-forming impact \citep{Canup+2001,Ruiz-Bonilla2021}. 

Previous studies that have addressed the smoothing of density discontinuities in a planetary context by modifying the SPH formulation have dealt either with those arising from contact between two different materials at the same pressure \citep{Woolfson2007, Reinhardt2019} or with that found at the surface of a planet \citep{Reinhardt+Stadel2017}.
In this paper we propose a novel, computationally cheap method to suppress the spurious numerical effects associated with density discontinuities, regardless of their context.

In~\cref{sec:methods}, we describe the basics of SPH~(\cref{sec:background}), the details of the density discontinuity problem we aim to solve and previous attempts~(\cref{sec:problems}), and finally our own method~(\cref{sec:corrections}).
In~\cref{sec:tests}, we perform some standard hydrodynamical tests~(\cref{sec:square}, \cref{sec:KH}), as well as testing a settling simulation of a planet~(\cref{sec:profiles}), and a variety of giant impacts between a proto-Earth and Theia with and without our method~(\cref{sec:impacts}) to search for differences.
Finally, conclusions are presented in~\cref{sec:conclusions}.

\section{Methods}
\label{sec:methods}

\subsection{Background theory} 
\label{sec:background} 

The fundamental idea of SPH is to reconstruct a density field from a set of discrete particles with masses $m_i$. The density $\rho$ at any point in space $\vec{r}$ is computed as a weighted sum of the masses of the neighbouring particles \citep{Monaghan1992} via
\begin{equation}
    \rho(\vec{r}) = \sum\limits_{j=1}^{N_{\rm ngb}} m_j W(\vec{r} - \vec{r}_j,h),
	\label{eq:SPH_density}
\end{equation}
where $W$ is the kernel function, which is a function of position, and $h$ is the smoothing length. We will be referring to the density of particle $i$ as $\rho_{i}\equiv \rho(\vec{r_i})$, where $\vec{r_i}$ is the position of particle $i$.

Once the densities are computed, we can use the intrinsic specific internal energy $u_i$ of a particle (or any other thermodynamic variable), and the equation of state assigned to it ($\mathrm{EoS},i$) to compute the pressure at the location of each particle via $P_{i}\equiv P_{\mathrm{EoS},i}(\rho_{i}, u_i)$.

At this point we can compute the hydrodynamical forces using
\begin{equation}
    F=\frac{\nabla P}{\rho}=\nabla\left(\frac{P}{\rho}\right) + \frac{P}{\rho^2}\nabla\rho.
	\label{eq:nhydro_1}
\end{equation}
Then, we can discretize the acceleration of each particle
\begin{equation}
    \vec{a_i} = -\sum\limits_{j=1}^{N_{\rm ngb}} m_j \left(\frac{P_{j}}{\rho_{j}^2} + \frac{P_{i}}{\rho_i^2}\right) \nabla_i W_{ij},
	\label{eq:nhydro_2}
\end{equation}
where $W_{ij}\equiv W(\vec{r}_i - \vec{r}_j,h_i)$. This formula was first derived using a discrete form of the action principle for an adiabatic fluid. The rate of change in internal energy for particle $i$ can be expressed as
\begin{equation}
    \frac{du_i}{dt} = \frac{1}{2}\sum\limits_{j=1}^{N_{\rm ngb}} m_j \left( \dfrac{P_{j}}{\rho_{j}^2} + \dfrac{P_i}{\rho_i^2} \right) \vec{v}_{ij} \cdot \nabla_i W_{ij},
	\label{eq:nhydro_3}
\end{equation}
where $\vec{v}_{ij}=\vec{v}_i - \vec{v}_j$. We will be referring to these two formulae above as the standard SPH equations of motion.

This is not the only choice of discretization that can be used. We will now briefly summarize the geometric density average force (GDF) method \citep{Wadsley2017}, ignoring artificial viscosity terms:
\begin{equation}
    \vec{a_i} = -\sum\limits_{j=1}^{N_{\rm ngb}} m_j \left(\frac{P_i + P_j}{\rho_i\rho_j}\right) \bar{\nabla_i W_{ij}},
    \label{eq:GDFhydro_1}
\end{equation}
\begin{equation}
    \frac{du_i}{dt} = \sum\limits_{j=1}^{N_{\rm ngb}} m_j \left(\frac{P_i}{\rho_i\rho_j}\right) \vec{v}_{ij} \cdot \bar{\nabla_i W_{ij}}.
	\label{eq:GDFhydro_2}
\end{equation}
It is worth noting that these equations come from a general form of \cref{eq:nhydro_1} presented already by \cite{Monaghan1992},
\begin{equation}
    \frac{\nabla P}{\rho}=\frac{P}{\rho^{\sigma}}\nabla\left(\frac{1}{\rho^{1-\sigma}}\right) + \frac{1}{\rho^{2-\sigma}}\nabla\left(\frac{P}{\rho^{\sigma-1}}\right),
\end{equation}
with $\sigma=1$. This choice was made in order to minimize errors in the vicinity of strong density gradients.
In addition to this choice, the GDF method also requires a symmetric gradient of the kernel, in order to have symmetrized force terms, namely
\begin{equation}
\begin{split}
    \bar{\nabla_i W_{ij}} &= \frac{1}{2}f_i\nabla_i W(\vec{r}_{ij}, h_j) + \frac{1}{2}f_j\nabla_j W(\vec{r}_{ij}, h_j),\\
	\label{eq:GDFhydro_3}
\end{split}
\end{equation}
where
\begin{equation}
\begin{split}
    f_i &= \sum\limits_{j=1}^{N_{\rm ngb}} \frac{m_j}{\rho_i}r_{ij}^2 W'\left(\frac{r_{ij}}{h_i}\right) \Bigg/ \sum\limits_{j=1}^{N_{\rm ngb}}\frac{m_j}{\rho_j}r_{ij}^2 W'\left(\frac{r_{ij}}{h_i}\right),
	\label{eq:GDFhydro_4}
\end{split}
\end{equation}
and ${r}_{ij}=\left|\vec{r}_i - \vec{r}_j\right|$, $W(\vec{r}_i - \vec{r}_j, h_i)=\frac{1}{h_i^3} W \left(\frac{r_{ij}}{h_i}\right)$, $W'(q)=\frac{1}{q}\frac{dW}{dq}$.\\
This formulation of SPH minimizes surface tension effects in multiphase flows, which can be a desirable feature for planetary SPH simulations where the mixing between materials is key to some problems. 

\cite{Wadsley2017} chose to use \cite{Wendland1995} kernels over the traditional cubic spline kernel \citep{Monaghan1992} because they don't suffer from the pairing instability \citep{Dehnen+Aly2012}. However the GDF method itself does not require a specific kernel. Hence we will use the traditional cubic spline kernel in this work when comparing different flavours of SPH, for simplicity.

\begin{figure}
	\includegraphics[width=\columnwidth]{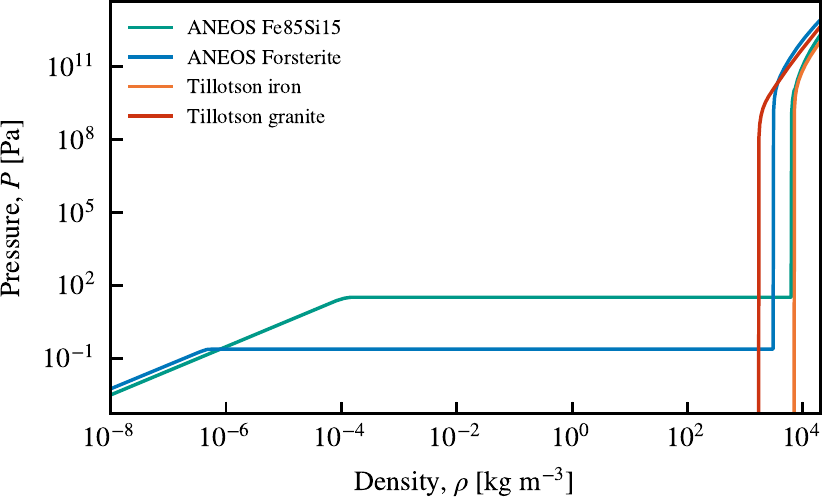}
	\vspace{-1em}
    \caption{Pressure as a function of density at a fixed temperature, $T=2000$~K, for different materials commonly used in planetary SPH simulations. Horizontal segments represent the phase transitions that are present only for more sophisticated equations of state like ANEOS. 
    }
    \label{plot_eos}
    \vspace{-1em}
\end{figure}

An important characteristic of planetary SPH simulations is the choice of materials, or in other words, equations of state (EoS). Each particle is labeled as being a particular material and, whenever needed, its equation of state is applied to compute its pressure. One well known and widely used option is the \cite{Tillotson1962} EoS. This analytical EoS was originally developed to model hypervelocity impacts, partly motivated by nuclear weapons research. Each material (e.g. iron, granite, etc.) is described by 10 parameters and 3 common analytical expressions describing a compressed or cold state, a hot and expanded state, and a hybrid state. As mentioned, this option is widely used for its simplicity but it has significant limitations. Materials described by the Tillotson EoS lack phase transitions, as well as not being suitable for giant impact simulations where vaporization plays an important role \citep{Stewart2020}. One example of a more modern approach is to use the ANEOS EoS \citep{Thompson1970, Melosh2007, Stewart2020}. This EoS model is described by the Helmholtz free energies for solid, liquid, vapor, plasma and mixed phases. It is capable of covering a large range of pressures, densities, and temperatures, which is important for simulating giant impacts between proto-planets. ANEOS EoS have over 40 input parameters, and multiple phase transitions are present for any material.
The presence or absence of phase transitions in the EoS used for the simulations, as shown in \autoref{plot_eos}, will have a key role when designing a method to solve our density discontinuity problem.

In this work we will use the open-source hydrodynamics and gravity code SWIFT \citep[SPH With Inter-dependent Fine-grained Tasking; \href{http://www.swiftsim.com}{www.swiftsim.com},][]{Schaller+2016, Kegerreis+2019}. SWIFT has been designed from scratch to run large simulations and scale well on shared/distributed-memory architectures. SWIFT runs over $30×$ times faster than Gadget-2 on representative cosmological problems \citep{Borrow+2018}, and has enabled planetary impact simulations with $100$-$1000$ times more particles than was previously typical. This speed is partly a result of SWIFT’s task-based approach to parallelism and domain decomposition for the gravity and SPH calculations \citep{Gonnet2015}.

\subsection{Problems in Planetary SPH}
\label{sec:problems} 

Given the definition of the density field in SPH (\cref{eq:SPH_density}), a direct consequence is that the density varies smoothly in space, which makes density discontinuities difficult to represent. However, differentiated planets in hydrostatic equilibrium can and should contain density discontinuities both where there is a change of material (e.g. the core to mantle boundary) and at the surface of the planet.

\begin{figure}
	\includegraphics[width=\columnwidth]{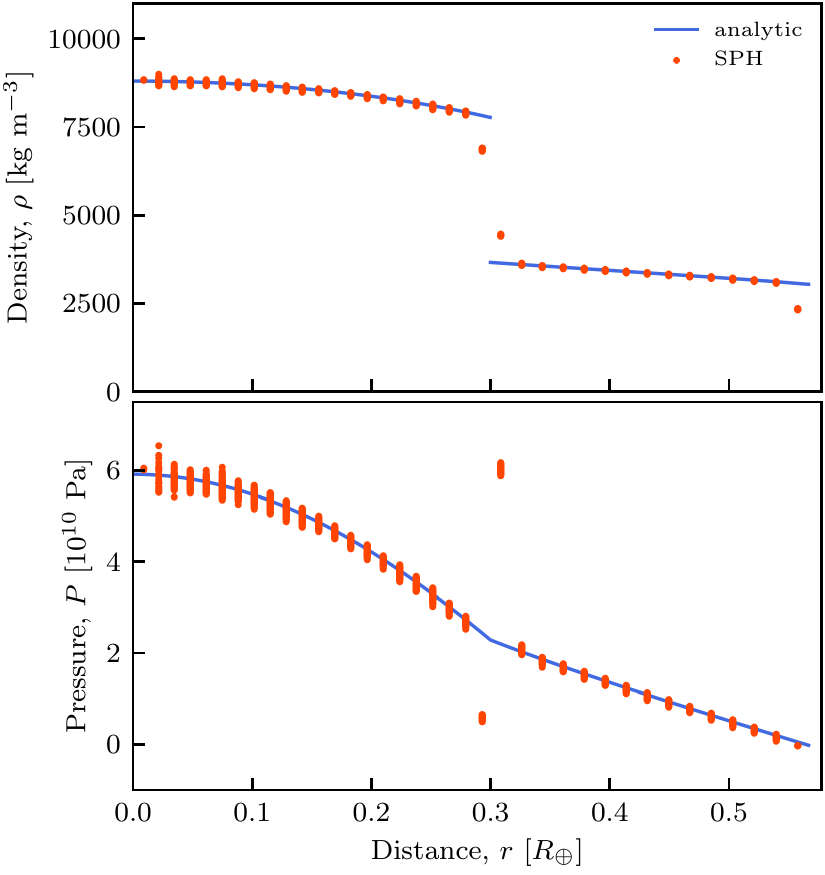}
	\vspace{-1em}
    \caption{Initial density (upper panel) and pressure (lower panel) profiles of a spherical Theia-like planet, $M=0.133$ M$_\oplus$, using standard SPH density calculations. The core:mantle mass ratio is 30:70,  and the temperature at the surface of the planet is $2000$ K with an adiabatic entropy profile. 10$^5$ particles are used to represent the planetary body. The smoothing in the density field introduces spurious pressures at the material boundary and, to a lesser extent, the edge of the planet.}
    \label{plot_problems}
    \vspace{-1em}
\end{figure}

When trying to represent a planet in SPH simulations, the smoothing of SPH particle densities across these discontinuities gives rise to well known problems \citep{Woolfson2007}, with poorly quantified consequences.
The incorrect pressures induced by the smoothed densities in these regions effectively create an artificial force that repels different material layers from each other. In the case of the free surface, particles in the outermost regions of the planet will have their densities underestimated. This will subsequently lead to underestimated pressures that will accelerate the system away from the desired equilibrium configuration.
We illustrate the initial problems using a Theia-like body, with mass $M=0.133$ M$_\oplus$, in~\autoref{plot_problems}. The analytical profile and particle placement for this, and all examples in this paper, have been produced by the open-source code WoMa \citep{Ruiz-Bonilla2021}, which uses the SEAGen method \citep{Kegerreis+2019} to make particle realisations of planets. Note that the artificial effective surface tension is equally present for both the standard and GDF flavours of SPH, because it arises from the definition of the density field, which is common to both methods.

\subsubsection{Density discontinuities between different material layers}
\label{subsec:diffmat}

Previous studies have attempted to address the issue of density discontinuities between different materials by changing the formulation of SPH \citep{Price2008, Hosono+2016b}. \cite{Woolfson2007} and later \cite{Reinhardt2019} proposed solutions based on computing correction factors, $f_{ij}$, for the SPH density, which was corrected via
\begin{equation}
\begin{split}
\label{eq:rein}
    \rho_i &= \sum\limits_{j=1}^{N_{\rm ngb}} f_{ij} m_j W_{ij},\\
    f_{ij} &= \frac{\rho_{\mathrm{EoS},i}(P, T)}{\rho_{\mathrm{EoS},j}(P, T)}.
\end{split}
\end{equation}
The density of particle $i$ is calculated with the inclusion of the
correction factor, which varies for each neighbouring particle
$j$. The correction factors represent the ratio of densities that particle $i$ would have at pressure $P$ and temperature $T$ if it were made from material $i$ versus material $j$. Thus, if neighbour $j$ is the same material as particle $i$, then $f_{ij}=f_{ji}=1$. \cite{Woolfson2007} applied this modification to equilibrium models of planets, where the temperature and pressure vary smoothly with radius. \cite{Reinhardt2019} suggested using the kernel-averaged temperature and pressure as better estimates to account for dynamical evolution of the system during a giant impact simulation.

This approach can reduce the problem, especially for stationary planets, but has a couple of inconvenient drawbacks.
The first is that, computationally, it requires three loops over all particles to compute the density, compared with the single loop used in the standard density definition: the first loop is used to compute the standard SPH density, the second one to compute the kernel averages of temperature and pressure, and the third one to recompute the density using \cref{eq:rein}. In addition to these, there is a final fourth loop to compute the hydrodynamical forces using \cref{eq:nhydro_2}, \cref{eq:nhydro_3} for standard SPH or \cref{eq:GDFhydro_1}, \cref{eq:GDFhydro_2} for GDF. The second and more serious downside appears when using more sophisticated equations of state like ANEOS. As shown in \autoref{plot_eos}, for the same temperature and pressure, two materials could have different densities by many orders of magnitude.
For example, at 10 Pa and 2000 K, the density ratio between Tillotson iron and granite is $0.2$, but between ANEOS Fe$_{85}$Si$_{15}$ and forsterite is $9.5\times 10^{7}$.
This occurs because one material (in this case Fe$_{85}$Si$_{15}$) has undergone vaporization whereas the other (forsterite) has not. When computing the final density using \cref{eq:rein}, the $f_{ij}$ factors could produce hugely unrealistic densities if particle $i$ has a significant number of neighbours $j$ of a different material. This issue would not only affect the density estimation of a few particles and hence the evolution of the system, but the high densities will also yield high pressures and hence forces that will dramatically decrease the value of the time step needed to continue evolving the simulation.

\subsubsection{The free surface problem}
\label{subsec:freesurface}

\cite{Reinhardt+Stadel2017} also proposed a solution, distinct from those described above, to the problem of the density discontinuity present at the surface of any planet. Their approach consisted of defining a statistic
\begin{equation}
\label{eq:rein_imb}
    f_i = \frac{\left|\sum\limits_{j=1}^{N_{\rm ngb}} \left(\vec{r}_j - \vec{r}_i\right)m_j W_{ij} \right|}{2 h_i \sum\limits_{j=1}^{N_{\rm ngb}} m_j W_{ij}}
\end{equation}
that is computed for every particle. The density of each particle is corrected by a factor that depends upon the value of this statistic, $f_i$, with the correction factor derived by assuming that the particle configuration involves a plane boundary between mass and vacuum in the kernel. This assumption may be appropriate to explain non-zero $f_i$ values during a simulation of a planet in hydrostatic equilibrium. However, there could be different scenarios where this is not the case during a planetary impact SPH simulation, for instance a satellite being tidally disrupted into some distorted geometry. Thus a more general approach to correcting densities near the boundary with a vacuum is desirable.

\subsection{Density corrections}
\label{sec:corrections}

Here we present our method to address both the material boundary and free surface problems at once.
In addition, this method is relatively computationally cheap since it only uses one extra loop over all particles compared with the standard SPH density computation.
We define a statistic that measures how afflicted a particle is by being close to a density discontinuity.
The densities of these particles are then corrected in a smooth way using that same statistic.

The method can be summarized as two steps: first we identify problematic particles, then we fix their densities.
Our first goal is to identify particles close to a material boundary or free surface. Our proposal, which is similar to that of \cite{Reinhardt+Stadel2017}, is
\begin{equation}
\label{eq:I}
\begin{split}
    I_i &= \alpha\frac{\left|\sum\limits_{j=1}^{N_{\rm ngb}} \kappa_{ij}\left(\vec{r}_j - \vec{r}_i\right)m_j W_{ij}\right|}{h_i \sum\limits_{j=1}^{N_{\rm ngb}} m_j W_{ij}},\\
    \kappa_{ij} &=
    \begin{cases}
      1 & \text{if $i$ and $j$ are the same material},\\
      -1 & \text{if $i$ and $j$ are different material}.\\
    \end{cases}   
\end{split}
\end{equation}
where $\alpha$ is a dimensionless parameter whose value we discuss later. We will refer to $I_i$ as the `imbalance statistic' for particle $i$.

Particles sitting in the middle of a perfectly regular grid of the same material particles will have an imbalance statistic equal to zero, and this will be approximately the case for most of the particles in our initial planet in hydrostatic equilibrium. Particles sitting at the surface of a planet will have about half of their kernel filled with particles of the same material and the other half empty. Their imbalance statistics should be somewhat greater than zero and, for the choice of $\alpha$ we describe in due course, they will be of order unity. Similarly, for particles placed at the boundaries between two materials, one half of their kernel is full of particles of the same material whereas the other hemisphere is full of particles of a different material; hence the inclusion of the minus sign in $\kappa_{ij}$ to account for the contributions from particles of the other material and produce a comparable unity-order value for $I_i$.

Now that we have defined the imbalance statistic that locates the problematic particles, we need to correct their densities. 
First, we compute the standard SPH density using \cref{eq:SPH_density}, and the pressure, $P_i$, and temperature, $T_i$, for every particle  using their corresponding equation of state with their density, $\rho_i$, and specific internal energy, $u_i$, which is used in the hydrodynamical simulation rather than the temperature.
Then, assuming that pressure and temperature vary smoothly on the scale of the smoothing length everywhere within the simulation, we compute, for every particle, estimated values for their temperature and pressure via

\begin{equation}
    \bar{T}_i = \frac{\sum\limits_{j=1}^{N_{\rm ngb}} T_j\mathrm{e}^{-I_j^2}W_{ij}}{\sum\limits_{j=1}^{N_{\rm ngb}} \mathrm{e}^{-I_j^2}W_{ij}},\hspace{2mm}\bar{P}_i = \frac{\sum\limits_{j=1}^{N_{\rm ngb}} P_j\mathrm{e}^{-I_j^2}W_{ij}}{\sum\limits_{j=1}^{N_{\rm ngb}} \mathrm{e}^{-I_j^2}W_{ij}}.
    \label{eq:TPbar}
\end{equation}
These estimates represent pressures and temperatures averaged over neighbouring particles, weighted to favour nearby neighbours with low imbalance statistics.
Recall that low imbalance statistic particles typically have neighbours sitting in regular grids, so those particles should be away from sharp density discontinuities and thus in regions where pressures and temperatures are computed accurately. 
In addition, we would like to have a smooth transition between the standard SPH computation and our modified one, such that the modification is only used when needed and without any sudden transitions. Hence we can define a pressure, $\widetilde{P}_i$, and temperature, $\widetilde{T}_i$, for every particle as
\begin{equation}
\begin{split}
    \widetilde{P}_i &= \mathrm{e}^{-I_i^2}P_i + (1 - \mathrm{e}^{-I_i^2})\bar{P}_i,\\
    \widetilde{T}_i &= \mathrm{e}^{-I_i^2}T_i + (1 - \mathrm{e}^{-I_i^2})\bar{T}_i,
    \label{eq:TP}
\end{split}
\end{equation}
such that the more problematic a particle is (the higher the imbalance statistic) the greater the contribution from the modified estimate.
Now that we have estimated a corrected pressure and temperature for every particle, we use the corresponding equation of state to infer a corrected density for every particle via
\begin{equation}
    \widetilde{\rho}_i = \rho_{\mathrm{EoS}, i}(\widetilde{T}_i, \widetilde{P}_i).
    \label{eq:rho_tilde}
\end{equation}
Finally, we compute a particle pressure based upon this corrected density and the unaffected specific internal energy, using $P_i = P_{\mathrm{EoS}, i}(\widetilde{\rho}_i, u_i)$. $\widetilde{\rho}_i$ and $P_i$ are the values that are used in the equations of motion.

\begin{figure}
	\includegraphics[width=\columnwidth]{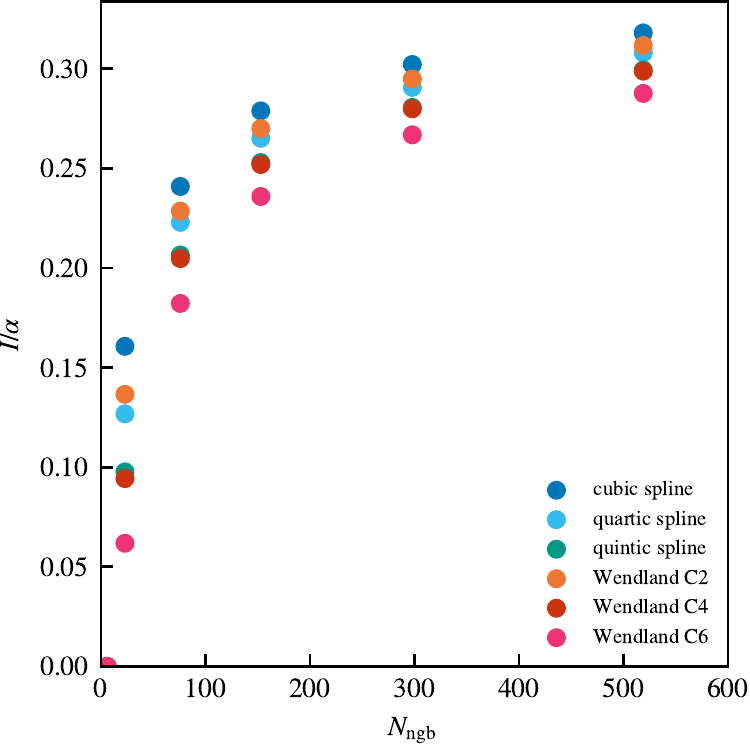}
	\vspace{-1em}
    \caption{The scaled imbalance statistic, $I/\alpha$ (\cref{eq:I}), calculated at the centre of a sphere, only half of which is filled by particles in a regular cubic grid, as a function of the numbers of neighbours.
    This value is computed to normalize the value of the imbalance statistic so that it effects the first shell of particles at a density discontinuity, but not the rest.
    Different colours represent a variety of kernels, as detailed in the legend.}
    
    \label{plot_alpha}
    \vspace{-1em}
\end{figure}

We determine the value of $\alpha$ with the following condition: a particle with a kernel that is half full of particles of the same material organized in a regular grid, and with the other half empty must have imbalance statistic equal to $1.5$.
We have chosen this value empirically, since an imbalance statistic of $1$ has very little effect on the densities of particles one shell away from a different material, and a value of $2$ significantly affects particles two shells away from a different material.
This ensures particles at the surface of the planet and the material boundaries will have big enough imbalance statistics for the method to have a significant impact, without overcorrecting. This parameter may need to be adjusted if the kernel and/or the resolution parameter $\eta$ (i.e. the number of neighbours within the kernel) is changed.
\autoref{plot_alpha} shows the reduced imbalance statistic, $I/\alpha$, for different kernels and numbers of neighbours. The minimum number of particles used to compute $I/\alpha$ is 6 ($(0,0,0)$, $(\pm 1,0,0)$, $(0,\pm 1,0)$, and $(0,0,-1)$). By making that grid finer we can compute it with 23, 76, 153, 298, and 519 particles. Finally we interpolate linearly to obtain the value of $\alpha$ that yields $I=3/2$, depending on the desired number of neighbours. 

\begin{figure}
	\includegraphics[width=\columnwidth]{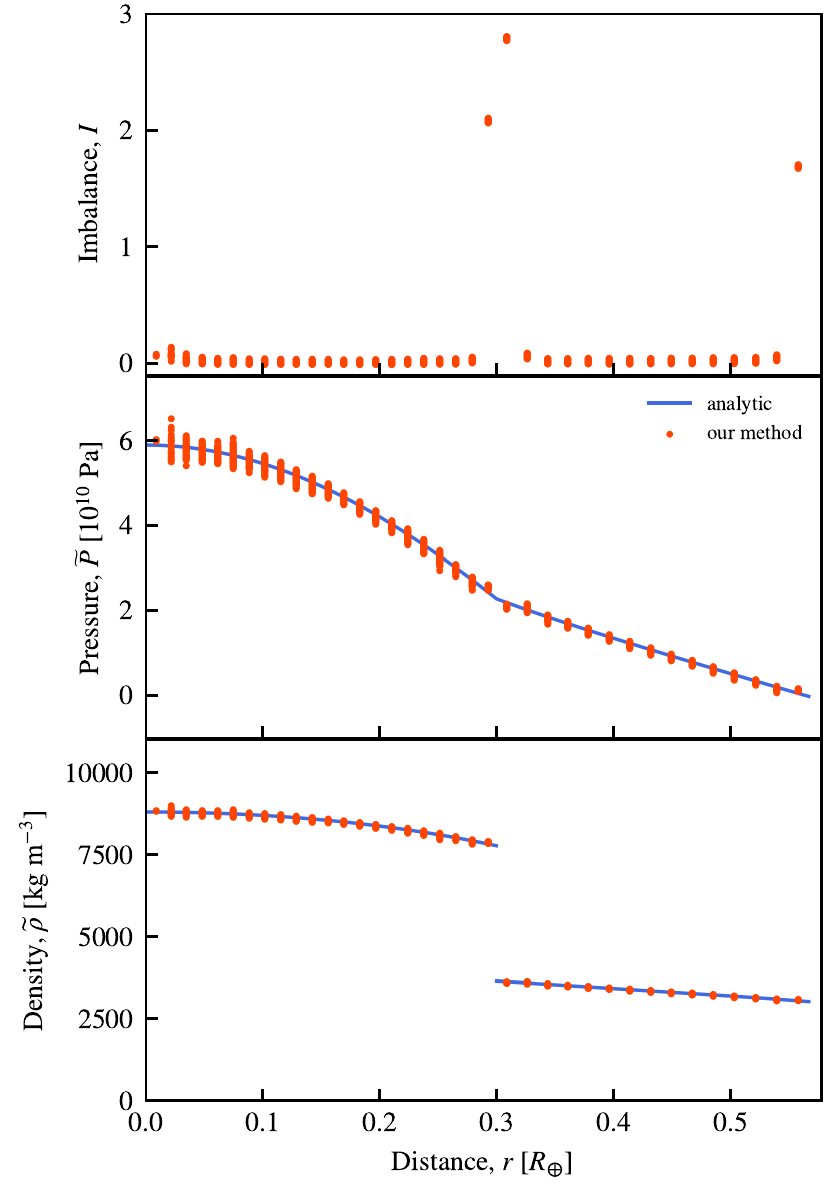}
	\vspace{-1em}
    \caption{Imbalance statistic, $I$, estimated pressure from \cref{eq:TP}, $\widetilde{P}$, and corrected density, $\widetilde{\rho}$, for the same spherical Theia-like planet, $M=0.133$~M$_\oplus$, used in \autoref{plot_problems}.}
    \label{plot_imb_P_rho}
    \vspace{-1em}
\end{figure}

This method only uses two loops over the neighbours of all particles. The first loop is used to compute the standard SPH density, pressure, and temperature, as well as the imbalance statistic for every particle; the second loop is used to evaluate \cref{eq:TPbar}, which leads to the corrected density, $\widetilde{\rho}_i$, and final pressure.
To illustrate how our method works, \autoref{plot_imb_P_rho} shows the imbalance statistic, the intermediate estimate of the pressure, and the final corrected density, for the same Theia-like planet that was shown in \autoref{plot_problems}.
The imbalance statistic targets the right particles and the weighted and smoothed pressure estimate erases the pressure jump that is present for standard SPH. Finally, using the equation of state, we compute corrected densities that have values close to the analytical solution.

\begin{figure*}
	\includegraphics[width=\textwidth]{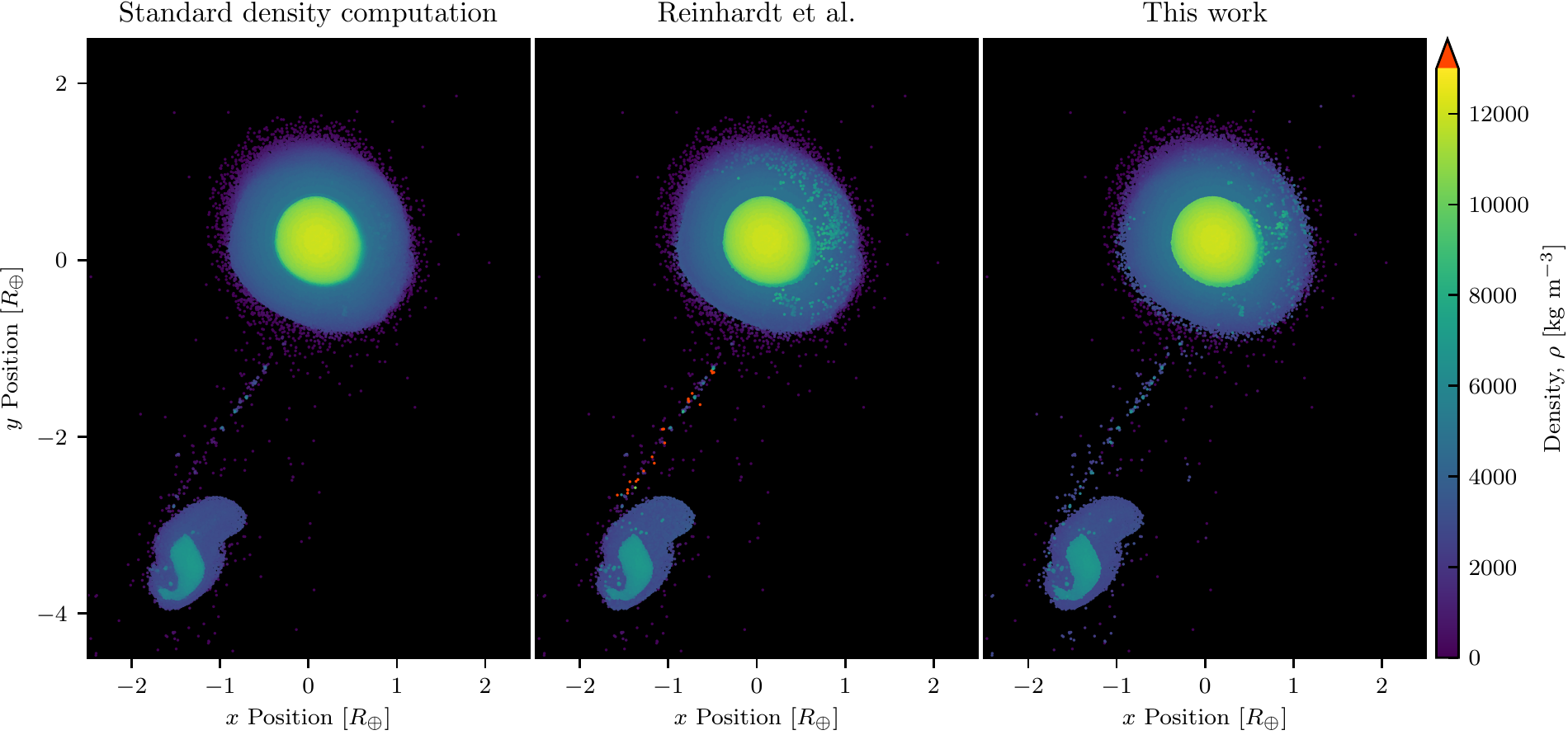}
	\vspace{-1em}
    \caption{Different methods for computing the density for the exact same particle configuration, a mid-collision snapshot of a canonical impact using ANEOS materials. Higher density particles are plotted on top of lower density ones. Red particles are at least one order of magnitude higher in density than the maximum shown by the colour bar.}
    \label{plot_same_snap_comparison}
    \vspace{-1em}
\end{figure*}

\autoref{plot_same_snap_comparison} demonstrates the different calculations of density produced for one (identical) snapshot of a simulation by the three methods: standard SPH (\cref{eq:SPH_density}), the \cite{Reinhardt+Stadel2017} method described in \cref{subsec:diffmat}, and our method described above. The exact same particle configuration is used in all cases, a mid-collision snapshot of a giant impact between a proto-Earth and Theia, using ANEOS materials. The different densities produced by the three methods would lead to different subsequent evolution of these cases, if they were evolved forward from this common starting point using the different methods.
The distinction between core and mantle material in the proto-Earth is quite diffuse in the standard SPH computation compared with the other two methods.
The \cite{Reinhardt+Stadel2017} method yields particles with densities over $10^5$ kg~m$^{-3}$, highlighted in red on the figure. 
This is due to the problem described in \cref{sec:problems} when using equations of state with phase boundaries like ANEOS. Finally, the iron particles scattered within the mantle of the proto-Earth are assigned significantly higher densities using the \cite{Reinhardt+Stadel2017} method or our method than for standard SPH, and thus are more clearly visible in the figure.
This is because their method acts whenever a particle has a neighbour of a different material, whereas in our method having a regular grid of particles of random materials will yield a density identical to the standard SPH one, up to the noise in the particle distribution for each material.
Hence our method produces high densities for iron particles in the mantle when they have other iron particles as neighbours, but not when they are just surrounded by granite particles.

\section{Tests and examples}
\label{sec:tests}

Up to this point we have been using static distributions of particles to compare different ways to compute the density field in SPH simulations. Now we will perform some dynamic tests and example simulations combining these density estimators with the two different equations of motion that were presented in \cref{sec:background}, standard SPH and GDF SPH \citep{Wadsley2017}. 

\subsection{2D Square Test}
\label{sec:square} 

One of the most common tests of contact discontinuities is the square test \citep{Saitoh+Makino2013}. A 2D box of a certain material and density is surrounded by a medium of the same or different material at a different density in pressure equilibrium. If the code does not capture the density discontinuity correctly, then the pressure at the material interphase becomes discontinuous. This creates an artificial tension, similar to that shown in \autoref{plot_problems}, which effectively acts to round the corners of the box.

For this test we use \cite{Tillotson1962} materials, which are often used in planetary SPH simulations.
The central square contains iron whereas the surroundings are composed of granite. The side length of the simulation box is $l_x=l_y=0.5$ R$_\oplus$, and the depth of the box is $l_z=0.001$ R$_\oplus$. This small thickness, together with periodic boundary conditions, allows densities to be computed in 3D despite particles being confined to 2D.
The setup is designed such that the pressure everywhere is $10^{10}$ Pa and the temperature is $1000$ K. These constraints dictate the mass (i.e. density) and internal energy of each particle, and result in a density jump from $\rho_{\rm granite}=3251$ kg~m$^{-3}$ to $\rho_{\rm iron}=7980$ kg~m$^{-3}$. Iron particles, located in the inner square of side length $l_x/2$, are given a larger mass than granite ones such that the $2^{14}$ total particles in our simulations can be placed onto a regular square grid. Each simulation is evolved until 100 ks, which corresponds to roughly 300 sound crossing times of the central iron square.

\begin{figure}
	\includegraphics[width=\columnwidth]{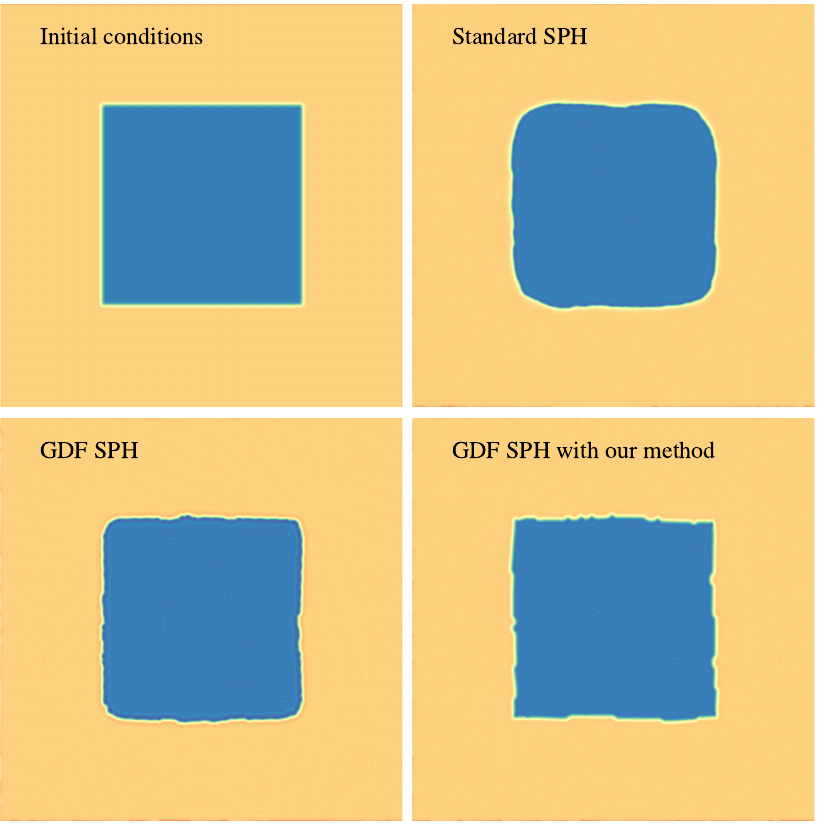}
	\vspace{-1em}
    \caption{2D Square test for 3 different flavours of SPH: standard SPH, GDF SPH, and GDF SPH with our method. Blue and orange represent higher and lower density values respectively. All panels, except for the initial conditions, show the result of the test after $100$ ks, which corresponds to $\sim$300 sound crossing times of the central iron square.}
    \label{plot_square_comparison}
    \vspace{-1em}
\end{figure}

We tested 4 different flavours of SPH: standard SPH; standard SPH with our method for improving densities; GDF (geometric density average force) SPH \citep{Wadsley2017}; and GDF SPH with our method.
Adding our method on top of the standard SPH equations of motion has little effect on the overall evolution, so we will not discuss this combination for any of the tests in this paper. \autoref{plot_square_comparison} shows the initial conditions and the result for the remaining three flavours of SPH following 100 ks of evolution.
Relative to standard SPH, GDF SPH better maintains the shape of the central square, and our method further improves the sharpness of the corners. 


\subsection{2D Kelvin-Helmholtz Test}
\label{sec:KH} 

The Kelvin-Helmholtz test is a common way to determine how well methods capture the instability that arises between adjacent fluids moving with different velocities.
This test does not have a known analytical solution, so we compare with different hydrodynamic codes, how our method handles the Kelvin–Helmholtz instability.

For this experiment we used the same box dimensions and materials as in the square test (\cref{sec:square}). The pressure and temperature throughout the box were set to $10^{10}$ Pa and $1000$ K, and the number of particles on the x-axis for the low density material was $N=256$. We chose all particles to have the same mass, so the denser layer contained $\sim$2.5 times more particles. The central strip of iron particles were given an initial $x$ velocity of $v_x=1000$~m~s$^{-1}$, whereas the granite particles had $v_x=-1000$~m~s$^{-1}$. An initial perturbation in the $y$ velocity, $v_y=20\sin(4\pi x/R_\oplus)$~m~s$^{-1}$, was introduced in order to seed the instability. 

\begin{figure}
	\includegraphics[width=\columnwidth]{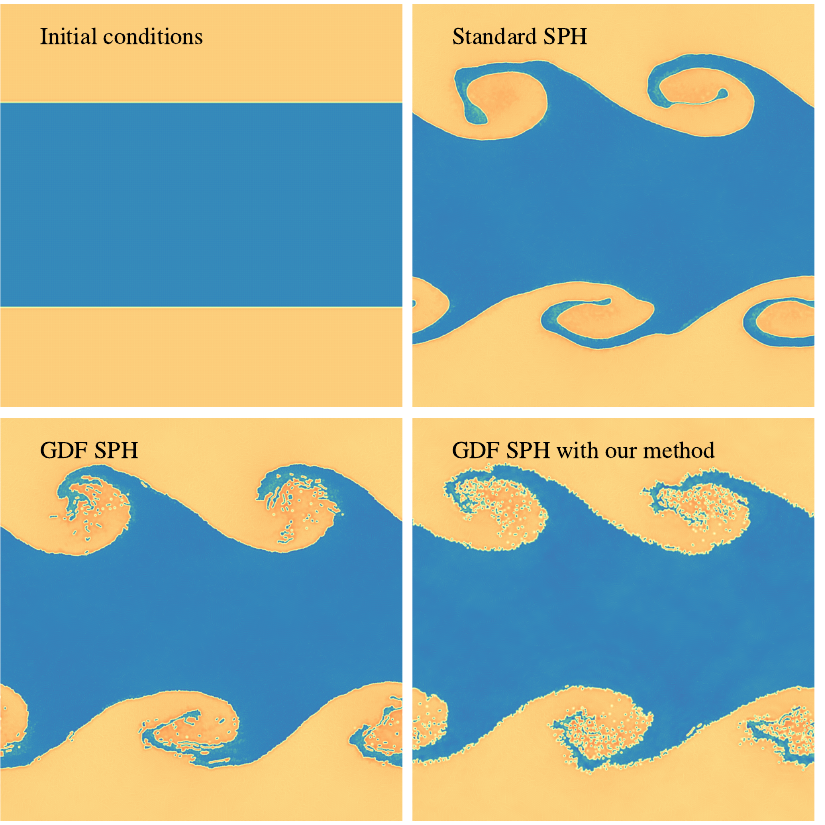}
	\vspace{-1em}
    \caption{2D Kelvin-Helmholtz test for 3 different flavours of SPH: standard SPH, GDF SPH, and GDF SPH with our method. Blue and orange represent Tillotson iron and granite respectively. All panels except that for the initial conditions show the result of the test after $2500$ s.}
    \label{plot_KH_comparison}
    \vspace{-1em}
\end{figure}

We observe that all three flavours of SPH create roll-like structures after $2500$~s, as shown in \autoref{plot_KH_comparison}. At that time a particle travelling at the original $x$ velocity will have traversed just over three quarters of the box length.
We can see clear qualitative differences between all three flavours of simulation and, relative to the other two cases, GDF SPH with our method shows enhanced mixing between different material particles at the end of the swirls. 

We tested the numerical convergence for a range of resolutions up to $N=2048$ and calculated the time evolution of the maximum $y$-direction kinetic energy density and the amplitude of the $y$-velocity mode of the instability. All three flavours of simulation produced very similar shapes to the SPH code results shown by \cite{mcnally+2012}. 
Using the smoothed $y$ velocity field values, rather than the individual SPH particle values, made the maximum $y$-direction kinetic energy density grow slightly slower, but in neither case was the evolution in this statistic comparable with that found for grid codes \citep{mcnally+2012}.
We conclude that our method does not converge significantly faster or slower than previous formulations of SPH for this particular test.

\subsection{Planetary profiles after settling simulations}
\label{sec:profiles} 
Prior to running a planetary giant impact simulation, a settling simulation is typically undertaken for every proto-planet. This is done in order to reduce any noise from the initial positions of the particles, as well as to obtain an object that is in hydrostatic equilibrium.
Ideally the SPH densities, which are computed using the positions and masses of the particles, should match those obtained from solving the hydrostatic equilibrium equation. However, particle placement algorithms always introduce some perturbations and, as is the focus of this study, density discontinuities are not well captured with the traditional SPH density computation.

We consider a proto-Earth like planet, $M=0.887$ $M_\oplus$, made of an ANEOS Fe$_{85}$Si$_{15}$ core and forsterite mantle \citep{Stewart2020} with a surface temperature of $T=2000$ K and adiabatic temperature profile. The core:mantle mass ratio is 30:70. We use approximately $10^6$ particles and let the simulation run for $20$ ks, which is many times the sound crossing time of the planet.

\begin{figure*}
	\includegraphics[width=\textwidth]{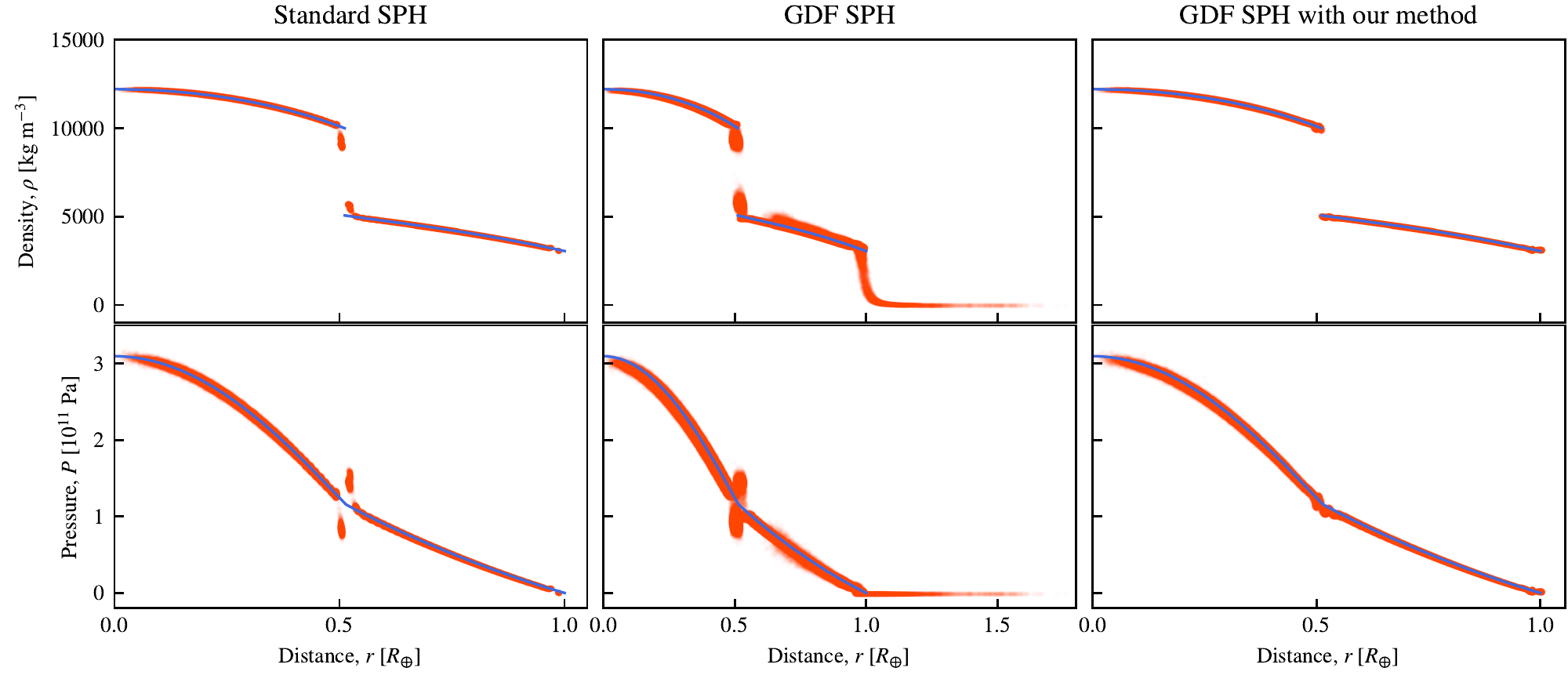}
	\vspace{-1em}
    \caption{Density (top row) and pressure (bottom row) profiles after 20 ks of a settling simulation of a proto-Earth in hydrostatic equilibrium, for different flavours of SPH (different columns). The blue line represents the analytical profile, and red dots represent particles in the simulation. The central column, showing the GDF SPH results, has an expanded radial scale to show the extent to which particles are leaking away.}
    \label{plot_settling}
    \vspace{-1em}
\end{figure*}

\autoref{plot_settling} shows the results for standard SPH, GDF SPH, and GDF SPH with our method. For standard SPH we see the evolved versions of the same issues that were highlighted for the initial particle arrangement in \autoref{plot_problems}. The density discontinuity in the material boundary is smoothed over, leading to a spurious jump in pressure across the boundary. Also, the underestimated density, and hence pressure, in the outermost shell of particles of the planet has decreased the radially outwards hydrodynamical force (see \cref{eq:nhydro_2}). Consequently, the planet has contracted to find its numerical equilibrium, leaving it slightly smaller than desired. 

For GDF SPH with the standard density computation, the situation is even worse because it does not even reach an equilibrium state. The density discontinuity is somewhat smoothed, although not quite as badly as in the standard SPH case. However, there is an additional problem whereby particles at the edge of the planet continually leak away; note the expanded horizontal scale for these panels.
The reason for this is the factor $f_i$ described in \cref{eq:GDFhydro_4}.
Consider the particles sitting at the edge of the main planet. Their neighbours are predominantly interior with higher densities. These outermost particles will have underestimated densities because of the exterior vacuum increasing their smoothing lengths. However, this problem will not affect the interior particles, which will have densities that more accurately reflect the input profile.
As a consequence, $f_{i}$ will be inappropriately large for the outermost particles, producing an outward hydrodynamical acceleration (\cref{eq:GDFhydro_1}) that exceeds the inward pull of gravity. Within two hours of simulation time, particles are already flying outwards. $f_i$ can reach values of the order of $100$, where the typical value should be around $1$, and this effect gradually peels off more layers from the outer edge of the planet.

GDF SPH with our method solves the problems mentioned above, as illustrated in \autoref{plot_settling}. Not only is the density discontinuity well represented, which means that there is no jump in the pressure profile between the core and mantle, but the outer boundary also closely matches the analytical profile, meaning that the planet has the intended radius.

\subsection{Giant impacts}
\label{sec:impacts} 

In this section we compare features that occur during giant impacts between planets for different flavours of SPH. In \cref{sec:profiles} we saw that GDF SPH needs to be accompanied by our method in order to have stable planets, therefore we will just consider standard SPH, and GDF SPH with our method.

We use the proto-Earth and Theia described in \cref{sec:profiles} and \cref{sec:problems}, respectively, increase the number of particles by a factor of $10$, and collide them with a range of impact angles and velocities. The total number of particles in our simulations is approximately $10^7$, with all particles having the same mass. 

We run three different impact scenarios, varying the angle of impact $\beta$ and the impact velocity at contact $v_{\rm c}$: a `canonical' impact ($\beta=45^{\circ}$, $v_{\rm c}=1~v_{\rm esc}$), a faster low-angle impact ($\beta=15^{\circ}$, $v_{\rm c}=2~v_{\rm esc}$), and a hit-and-run grazing impact ($\beta=65^{\circ}$, $v_{\rm c}=1.5~v_{\rm esc}$). The mutual escape velocity of the system is $v_{\rm esc}=9026$~m~s$^{-1}$. Each impact is run four times using different random reorientations of the particle realizations of the planets. This provides an estimate of the stochastic noise and allows us more confidently to ascribe any observed differences to the different flavour of SPH being used. Depending upon just how chaotic the impact and its aftermath are, this can be an important consideration (Kegerreis et al. 2022 in prep).

We have now resolved the density discontinuity issues we previously had, and can thus be confident that these are not causing big unknown errors.
Beyond the broad similarities, we observe some key differences between both flavours of SPH. These are common to all of the randomly reoriented resimulations and so appear to be robust differences between the standard SPH case and that with GDF SPH plus our method.

Changing between the two SPH flavours leads to significant differences in the distribution of post-impact iron in our low-angle collisions.
The mass-fraction of iron in the debris beyond $3\,R_\oplus$ is $\sim 8\%$ using our flavour of SPH. This is about three times higher than the corresponding value for standard SPH.
Within the final planet, the transition region between core and mantle, defined as the region where the relative iron content as a function of distance drops from $99\%$ to $10\%$, is $0.12~R_{\oplus}$ for standard SPH and $0.41~R_{\oplus}$ for GDF SPH with our method. This demonstrates how mixing of materials can increase if the spurious boundary pressure gradients associated with standard SPH are suppressed using GDF SPH and our method.
For both the canonical-like and hit-and-run impacts, the core of the target is barely disrupted by the impactor and the distribution of post-impact iron is insensitive to the flavour of SPH used.


In the initial conditions for our iron and rock bodies, the fraction of particles that have their densities badly mis-estimated by standard SPH as a result of their proximity either to a material boundary or the edge of the planet is $\sim14\%$ for a 10$^5$-particle realisation. This drops to a still large $\sim7\%$ with 10$^6$ particles, and $\sim 3\%$ for 10$^7$ particles.
The fraction of particles that at some point during an impact simulation have $I>1.5$, the value at the surface of a planet, is much larger. For 10$^7$-particle simulations, this fraction is $10\%$, $30\%$, and $70\%$ for hit-and-run, `canonical', and low-angle impacts respectively.
The spurious density is often sufficiently wrong that the particle will be translated across a phase boundary in its EoS. In addition to producing spurious pressure, this will complicate efforts to track the thermal evolution of the material, both during the impact simulation and when providing inputs for subsequent long-timescale thermal evolution codes. This is relevant for material in the target and the resulting debris, be it a diffuse disk or in coherent clumps \citep{Ruiz-Bonilla2021}. The combination of GDF SPH with the method we have described here practically eliminates these problems that are present in standard SPH approaches, opening up the opportunity to use SPH planetary giant impact simulations for such studies reliably.

\section{Conclusions}
\label{sec:conclusions}

We have presented a novel method to compute the density field in smoothed particle hydrodynamics (SPH) simulations with particular reference to scenarios of planetary giant impacts. It solves problems that arise in SPH for systems with sharp density discontinuities between different materials and between any material and a vacuum, with low computational cost.
We combine this method with the geometric density average force (GDF) SPH \citep{Wadsley2017} equations of motion because of their treatment that minimizes spurious numerical surface tension effects in multiphase flows. An implementation of our method is publicly available as an option in the open-source code SWIFT \citep{Schaller+2016}.

This new method produces improved performance in the 2D square test with a better maintained square shape, and enhanced mixing between different material particles in the 2D Kelvin-Helmholtz test.
Simulations of impacts between a proto-Earth and Theia, where the core of the Earth has been highly disrupted by the impactor, reveal a partially diffused iron core and a higher mass of iron in the debris disk.
This method also prevents smoothed densities from placing particles into inappropriate places in their material phase diagram. As a consequence, the thermodynamic evolution of material can be tracked more realistically throughout an impact and its aftermath.

\section*{Acknowledgements}


SRB is supported by a PhD Studentship from the Durham Centre for Doctoral Training in Data Intensive Science, funded by the UK Science and Technology Facilities Council (STFC, ST/P006744/1) and Durham University. VRE and RJM acknowledge support from the STFC grant ST/P000541/1. JAK acknowledges support from a NASA Postdoctoral Program Fellowship and STFC grants ST/N001494/1 and ST/T002565/1. TDS acknowledge support from the STFC grants ST/T506047/1 and ST/V506643/1. LFAT acknowledges support from NASA Emerging Worlds  Program award 80NSSC18K0499. 

This work used the DiRAC@Durham facility managed by the Institute for Computational Cosmology on behalf of the STFC DiRAC HPC Facility (www.dirac.ac.uk). This equipment was funded by BIS National E-infrastructure capital grant ST/K00042X/1, STFC capital grants ST/H008519/1 and ST/K00087X/1, STFC DiRAC Operations grant ST/K003267/1 and Durham University.

The research in this paper made use of the SWIFT open-source simulation code \citep[\href{http://www.swiftsim.com}{www.swiftsim.com},][]{Schaller+2016} version 0.9.0.

\section*{Data availability}
The data underlying this article will be shared on reasonable request to the corresponding author.




\bibliographystyle{mnras}
\bibliography{paper}




\appendix




\bsp	
\label{lastpage}
\end{document}